\documentclass[twocolumn,showpacs,preprintnumbers,amssymb]{revtex4}
\usepackage{graphicx}
\usepackage{dcolumn}
\usepackage{bm}

\begin{document}
\title
{Spin-Orbital Fluctuations and a large mass enhancement in LiV$_2$O$_4$}
\author{Yasufumi Yamashita$^{1,2}$ and Kazuo Ueda$^{1,3}$}
\affiliation{
$^1$Institute for Solid State Physics, University of Tokyo, Kashiwa, 
Chiba 277-8581, Japan\\
$^2$Department of Physics, University of Cincinnati, Cincinnati OH 45221\\
$^3$Advanced Science Research Center, Japan Atomic Energy Research
Institute, Tokai, Ibaraki 319-1195, Japan}
\date{\today}
\begin{abstract}
We present a scenario that the multi-component fluctuations, 
especially those of the spin-orbital coupled modes, 
lead to the mass enhancement observed in ${\rm LiV_2O_4}$.
This phenomena is possible because all these modes are 
fluctuating due to the geometrical frustration. 
To illustrate this mechanism, the $t_{2g}$-orbital Hubbard model 
on the pyrochlore lattice is studied based on the random-phase
approximation. We derive the generalized susceptibility in the SU(6)
spin-orbital space and calculate the free energy by using a
coupling-constant integration. 
The estimated specific heat coefficient is of the correct order of
magnitude to explain the experiment.
\end{abstract}
\maketitle

\section{introduction}
The metallic spinel ${\rm LiV_2O_4}$ is a unique 3$d$ heavy-fermion
(HF) compound\cite{kondo}. Its $\gamma$ coefficient amounts to $420$
mJ/mol$\cdot$K$^2$ and, strikingly, 
the low-temperature properties below $T^*=20 \sim 30$ are quite similar to
those of the lanthanide or actinide HF compounds\cite{kondo,urano}.
Since the $d$ electrons, 1.5 per vanadium ion, 
are responsible for both transport and magnetism in ${\rm LiV_2O_4}$,
it is not trivial whether one can apply
a scenario analogous to the conventional HF mechanism.
Moreover, the large spatial extent of $d$ orbitals compared with
well-localized $f$ orbitals is also unfavorable for the electrons
to be treated as localized. In fact, some transport properties, such as
the steep increase of resistivity above $T^*$\cite{urano} and the 
pressure-induced metal-insulator transition\cite{urano2}, are different
from those of the typical $f$-electron HF compounds.

Apart from the HF behavior, geometrical frustration is
another specific feature of this system. 
The B site of the normal-spinel (${\rm AB_2O_4}$) is known to form the
network of corner-sharing tetrahedra (pyrochlore lattice). 
In most spinels, the geometric frustration is partially released by the
structural phase transition. As a result, the ordered ground state is realized
in the tetragonal phase, such as the possible charge ordering in ${\rm
Fe_3O_4}$ or the N\'{e}el ordering in 
${\rm ZnV_2O_4}$\cite{ueda}. 
In contrast, ${\rm LiV_2O_4}$ remains cubic
and no static magnetic order was observed down to 0.02 K.
The Curie-Weiss fit of the susceptibility ($100\le T\le 300$) 
results in an effective spin-1/2 moment per V with an antiferromagnetic
(AF) coupling ($\Theta_{\rm cw}=-37K$)\cite{urano}. 
AF long range order (LRO) will be 
suppressed by the geometric frustration and, indeed,
the inelastic neutron scattering experiment captures
the development of short-range AF correlations below 
$\Theta_{\rm cw}$\cite{lee}.
Interestingly enough, the large enhancement of $\gamma$ is shared by  
other geometrically frustrated systems with strong AF spin fluctuations, 
like ${\rm Y(Sc)Mn_2}$\cite{shiga} and $\beta$-Mn\cite{nakamura}. 
These facts seem to suggest that the formation of the heavy-mass
$d$ electron is related to the geometric frustration. 
In view of the geometric frustration, it should be noted that, 
already at the 10\% electron doping, the HF ground state is easily destroyed 
into the spin-glass phase\cite{ueda}. 

In ${\rm LiV_2O_4}$, the crystal filed around each V atom is nearly cubic 
with a slight trigonal distortion. 
The band-structure calculations indicate that 
the Fermi level lies within the cubic-$t_{2g}$ manifold 
with a bandwidth of about 2 eV\cite{anisimov, matsuno, eyert, singh}, which
is actually the superposition of the two components; 
the relatively narrow $A_{1g}$ ($\sim$ 1eV) and 
the broad $E_{g}$ orbital ($\sim$ 2 eV) of the $D_{3d}$ group.
Matsuno $et\,al.$ considered this trigonal split less important,
and stressed the importance of the geometric frustration and/or
orbital degeneracy\cite{matsuno}. On the other hand, 
the LDA+$U$ calculation by Anisimov $et\,al.$ suggested that
the correlation effect leads to the orbital polarization, where the
$A_{1g}$ orbital is occupied singly to form a localized spin-1/2 
moment and the rest 0.5 electron fills the $E_{g}$ band.
Thereby, they proposed the Kondo scenario
based on the effective Anderson impurity model\cite{anisimov}. 

Followed by these band-structure calculations, 
various theoretical models are proposed 
in connection with the microscopic origin of the 3$d$ 
HF behavior\cite{kusunose, flude, fujimoto, lacroix, hopkinson, laad}.
Especially, the role of the geometric frustration has been featured in 
recent studies. Motivated by the Curie-Weiss law around
room temperature, the formation of the localized $s=1/2$ is assumed
by several authors 
on the grounds of the orbital polarization\cite{lacroix, laad} 
or the local valence fluctuations\cite{hopkinson}. 
In these treatments the spin frustration is their main concern.

In this paper, we consider that the present system realizes the 
prototypical itinerant frustrated model and 
investigate the $t_{2g}$-band pyrochlore Hubbard model
with particular emphasis on orbital fluctuations.
Within the random phase approximation (RPA) scheme, 
all kinds of fluctuations are included simultaneously without prejudice.
In principle, geometric frustration prevents any kinds of LRO and
thus the large $\gamma$ would be expected by the resultant 
enhanced fluctuations. In this scenario,
not only the spin but also both of the orbital and spin-orbital 
fluctuations are responsible for the enhancement.
To our knowledge, this is the first theoretical attempt 
to attribute the heavy-mass $d$ electron directly to the orbital 
degrees of freedom.
The instability itself of the same model was discussed by 
Tsunetsugu\cite{tsunetsugu}.
We have explicitly calculated 
the $T$-linear coefficient of the imaginary part of the 
generalized susceptibility and 
found the large enhancement of $\gamma$
mainly due to the spin-orbital fluctuations.

\section{Model Hamiltonian }
\begin{figure}[t]
\includegraphics[width=86mm]{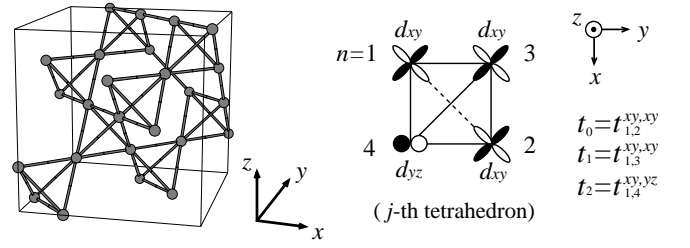}
\caption{The cubic unit cell of the pyrochlore lattice (left panel) and 
the three independent hoppings (right)}
\label{lattice}
\end{figure}

The pyrochlore lattice is a fcc-array of tetrahedra.
The cubic unit cell shown in Fig. \ref{lattice}
contains 16 lattice points covered by 4 tetrahedra and 
the primitive unit cell includes a single tetrahedron.
Therefore the $t_{2g}$ electrons are specified by the following 4 indices; 
the unit cell (denoted by $j=1,\cdots, 4L^3$), 
the sublattice ($n=1,\cdots, 4$), 
the orbital ($m=d_{xy}, d_{yz}, d_{zx}$), and the spin 
($\sigma=\uparrow/\downarrow$), where $L$ is the number of
the cubic unit cell along one direction.  
By using the standard notation of the multi-band Hubbard model,
the $t_{2g}$-band Hubbard model on the pyrochlore lattice
is given by ${\cal H}={\cal H}_0+{\cal H}_{\rm I}$;
\begin{eqnarray}
{\cal H}_0&=&\!\!\!\!\sum_{k\sigma mnm'n'}
{t}^{mm'}_{nn'k}c^{\dagger}_{kmn\sigma}c_{km'n'\sigma}\label{eq1}\\
{\cal H}_{\rm I}&=&\sum_{jn}\Bigg\{
U\sum_m n_{jmn\uparrow}n_{jmn\downarrow}
+U'\!\!\!\!\sum_{m>m' \sigma\sigma'}\!\!\!n_{jmn\sigma}n_{jm'n\sigma'}\nonumber\\
&&\hspace*{.6cm}-J\sum_{m\ne m'}
c^{\dagger}_{jmn\uparrow}c_{jmn\downarrow}
c^{\dagger}_{jm'n\downarrow}c_{jm'n\uparrow}\Bigg\}.\label{eq2}
\end{eqnarray}
Here $t^{mm'}_{nn'}$ is the transfer integral between the 
$m$ orbital at $n$ sublattice and the $m'$ orbital at $n'$ sublattice
when $(m,n)\ne (m',n')$.
In addition, the parameter of $t^{mm'}_{nn}(=\pm D/3)$ type is
also included to describe the trigonal split of $t_{2g}$, 
$E_{a_{1g}}-E_{e_g}=D$.
Since the nearest-neighbor (NN) hoppings already depend on 
the bond direction and the orbital symmetry
the NN tight-binding model would be sufficient for the present study.
By the symmetry relation, all the hopping matrices are
represented by the following three independent parameters,  
$t_0=t_{12}^{xy,xy}$, $t_1=t_{13}^{xy,xy}$, and $t_2=t_{14}^{xy,yz}$.
The $T_d$ symmetry of the unit cell, as well as the spin SU(2), are 
incorporated in the $12\times 12$ matrix ${\cal H}_0$, though here
we do not display the matrix explicitly.

The tight-binding fit of the LAPW band-structure calculation gives
the parameters in ${\cal H}_0$ to be $t_0=-0.281$ eV, $t_1=-t_2=0.076$
eV, and $D=0.140$ eV\cite{matsuno}. By applying the orthogonal transformation 
$c_{k\mu\sigma}=\sum_{\nu}U_{\mu\nu}(k)a_{k\nu\sigma}$, the single-electron
Hamiltonian is diagonalized into  ${\cal H}_0=\sum_{k\sigma\nu} 
\epsilon_{k\nu}a^{\dagger}_{k\nu\sigma}a_{k\nu\sigma}$, where
$\mu=(m,n)$ and $\nu$ are 12-dimensional indices.
The calculated $t_{2g}$-multiplet ($\epsilon_{k\nu}, \nu=1,\cdots, 12$) 
are shown in Fig. 3 along some typical symmetric lines. 
The almost flat bands near the Fermi level results in
high density of state (DOS) at the Fermi level.
Actually, the $\gamma$ coefficient estimated
from the DOS of the band-structure calculation is
relatively large;  $\gamma_{\rm band} \sim$ 17 mJ/mol$\cdot$K\cite{matsuno}. 
However, the experimentally observed value ($\gamma_{\rm exp}$)
is about 25 times larger than $\gamma_{\rm band}$. 
As a possible origin of this enhancement, in the following, 
we consider the effect of electron correlations from the weak-coupling limit.

\begin{figure}[t]\begin{center}
\includegraphics[width=86mm]{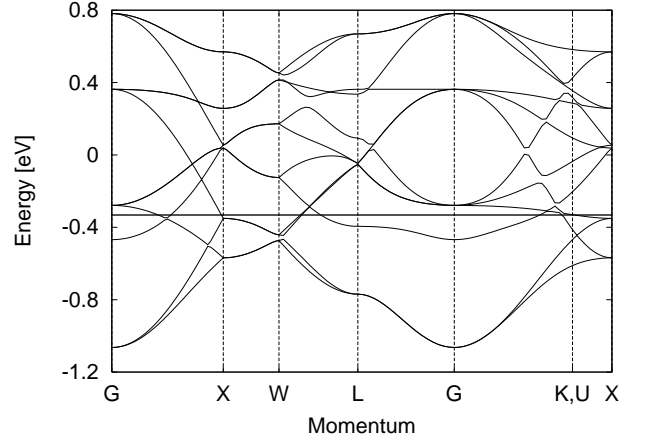}
\caption{Band structure of the $t_{2g}$ multiplet for the present parameter
along symmetric lines of the fcc Brillouin zone. The horizontal straight
line is the Fermi level for the quarter filling.}
\label{band}
\end{center}\end{figure}

\section{Free energy and RPA equations}

In order to describe various kinds of fluctuations concerning the $t_{2g}$
electrons in a simple way,
we introduce the thirty-five generators of the SU(6) group
$X^{\gamma} (\gamma =1,\cdots, 35$) and the identity operator $X^{0}$,
where the normalization condition is 
${\rm Tr\;} X^{\gamma}X^{\gamma'}=
\delta^{\gamma\gamma'}/2$.
The SU(6) generators are classified into the 
pure spin, the pure orbital and the spin-orbital coupled components.
Spin and orbital degrees of freedom are described by the generators of
SU(2) ($S^{\alpha}$; three dimensions) and 
SU(3) ($T^{\beta}$; eight dimensions) groups,
respectively, and the spin-orbital coupled modes are made of the
products of the two ($2\sqrt{3}S^{\alpha}T^{\beta}$; twenty four dimensions).
To be explicit, these operators are represented by 
using the electron creation and annihilation operators as follows,
\begin{eqnarray}
X^{0}_{jn}&=&\frac{1}{2\sqrt{3}}\sum_{\sigma m}
c^{\dagger}_{jmn\sigma}c_{jmn\sigma},\\
S^{\alpha}_{jn}&=& \frac{1}{2\sqrt{3}}
\sum_{\sigma\sigma'm}c^{\dagger}_{jmn\sigma}
\sigma^{\alpha}_{\sigma\sigma'}
c_{jmn\sigma'},\\
T^{\beta}_{jn}&=& \frac{1}{2\sqrt{2}}
\sum_{\sigma mm'}c^{\dagger}_{jmn\sigma}
\lambda^{\beta}_{mm'}
c_{jm'n\sigma},\\ 
2\sqrt{3}S^{\alpha}T^{\beta}_{jn}&=& \frac{1}{2\sqrt{2}}
\sum_{\sigma\sigma'mm'}c^{\dagger}_{jmn\sigma}
{\sigma}^{\alpha}_{\sigma\sigma'}\lambda^{\beta}_{mm'}
c_{jm'n\sigma'},
\end{eqnarray}
where $\vec{\sigma}$ and $\vec{\lambda}$ are the Pauli and 
the Gell-Mann matrices in the standard notations\cite{group}, respectively.
Note that the overall signs of non-diagonal operators for
different sublattices are selected to be consistent with the 
$T_d$ unit-cell symmetry.
Since the trigonal distortion in ${\cal H}_0$ violates the 
$O(3)$ orbital symmetry,
the usual representations of the $t_{2g}$ orbital by the
fictitious $\vec{l}=1$ and its higher moments
do not reflect the proper symmetry. 
Instead we construct the tensor products of the real $\vec{l}=2$ moment and
reduce them into the irreducible representations of $T_d$ 
in the cubic $t_{2g}$ subspace.
As a result, we find that $\vec{\lambda}$ is the proper set of basis,
which is classified into three irreducible representations;
${\mit\Gamma_5}$-dipolar moments ($\lambda_2,\lambda_5,\lambda_7$),
${\mit\Gamma_5}$-quadrupolar moments ($\lambda_1,\lambda_4,\lambda_6$), and
${\mit\Gamma_3}$-quadrupolar moments ($\lambda_3,\lambda_8$).

Next we transform the interaction part of Hamiltonian 
in terms of $X^{\gamma}$'s 
so that 
\begin{eqnarray}{\cal H}_{\rm I}=\sum_{jn}\sum_{\gamma=0}^{35}
a_{\gamma\gamma}X^{\gamma}_{jn}X^{\gamma}_{jn}\label{xx}\end{eqnarray}
except for a constant energy shift.
Here $a_{\gamma\gamma}$'s are a function of $U,U'$ and $J$.
Such transformation is always possible for any $U,U'$ and $J$, 
though the choice of $a_{\gamma\gamma}$ is not unique because of the spin
rotational invariance and/or the identity $n_{jmn\sigma}^2=n_{jmn\sigma}$.
By applying a coupling constant integration together with
Eq. (\ref{xx}), the Free energy for this system is given by
\begin{eqnarray}
\hspace*{-.5cm}F(U,U',J)&=&F(0)+
\frac{1}{2\pi}\int_{-\infty}^{\infty}d\omega
\coth{\frac{\beta\omega}{2}}\nonumber\\
&&\hspace*{-.8cm}\times\sum_{qn}\sum_{\gamma}a_{\gamma\gamma}
\int_0^1dx {\it Im}{\chi_{nn}^{\gamma\gamma}(q,\omega)}\Big|_{xU,xU',xJ},
\label{F}
\end{eqnarray}
where the momentum sum is taken all over the 1st Brillouin zone and
the generalized susceptibility per tetrahedron is defined by
\begin{eqnarray}
\chi_{nn'}^{\gamma\gamma'}\left(\vec{q},\omega \right)=\frac{i}{4L^3}
\!\int_0^{\infty}\hspace*{-.3cm}d\tau e^{i(\omega+i\delta)\tau}
\Big\langle\big[
X_{qn}^{\gamma}\left(\tau\right),X_{-qn'}^{\gamma'}\left(0\right)
\big]\Big\rangle.
\label{kai}
\end{eqnarray}

We apply the random phase approximation (RPA)
in the evaluation of the generalized susceptibility ($\chi$).
Including the charge degree of freedom ($X^{0}$), 
our $\chi$ is the $144\times 144$ 
matrix with the row index $n\gamma$ and the column $n'\gamma'$. 
Since we are concerned with the paramagnetic state, 
as suggested by the experiment, the SU(2) symmetry ensures that
$\chi$ is decomposed into the spin singlet and triplet sectors.
Each sector is represented by the charge-charge 
(denoted by $\chi_{nn'}^{\beta\beta'(c)}$) and the
$S^z$-$S^z$ susceptibilities ($\chi_{nn'}^{\beta\beta'(s)}$), respectively.
In these two $36\times 36$ matrices, the spin indices ($\alpha$) is 
reduced and the rest indices are the row index $n\beta$ and the column 
$n'\beta'$.
Here $\beta=0$ is defined to represent the totally symmetric orbital operator.
To sum up,
$\chi_{nn'}^{\beta\beta'(c)}$ describes the correlations
among the charge ($\beta=0$) and the orbital ($\beta \ge1$), while
$\chi_{nn'}^{\beta\beta'(s)}$ among the spin ($\beta=0$) and the
spin-orbital ($\beta \ge1$).
By using these notations, the RPA equations are given as follows,
\begin{eqnarray}
\chi_{nn'}^{\beta\beta'(c/s)}
=\chi_{nn'}^{\beta\beta'(0)}
+\sum_{n_1\beta_1}\chi_{nn_1}^{\beta\beta_1(0)}
2\Lambda_{\beta_1\beta_1}^{(c/s)}\chi_{n_1n'}^{\beta_1\beta'(c/s)},\label{rpa}
\end{eqnarray}
Since the interaction is local, the interaction matrices 
($\Lambda^{(c/s)}$) are always $n$-diagonal and independent of $n$.
Thus the rest indices are row $\beta$ and column $\beta'$.
Moreover the $9\times9$ matrices $\Lambda^{(c/s)}$ may have a simple
representation. 
This is because ${\cal H}_0$ has the $T_d$ symmetry and 
Eq. (\ref{rpa}) is invariant under any $T_d$ transformation.
Therefore, $\Lambda^{(c/s)}$'s must be  diagonal and these matrix
elements are the same within each irreducible representation,
see Table \ref{table1}. 
\begin{table}[tb]
\caption{\label{table1}The matrix elements of the diagonal
interaction matrices, $\Lambda_{\beta\beta'}^{(c/s)}$, for
each irreducible representation.}
\begin{ruledtabular}
\begin{tabular}{lcccc}
&$A_1 (\beta=0)$ &${\mit\Gamma_5} (1,4,6)$&
${\mit\Gamma_5} (2,5,7)$& ${\mit\Gamma_3} (3,8)$\\ \hline
$\Lambda^{(c)}_{\beta\beta}$& 
$-U-4U'+2J$ & $U'-2J$ & $U'-2J$ & $-U+2U'-J$\\ 
$\Lambda^{(s)}_{\beta\beta}$& 
$U+2J$ & $U'$ & $U'$ & $U-J$
\end{tabular}
\end{ruledtabular}
\end{table}

$\chi_{nn'}^{\beta\beta'(0)}$ in Eq. (\ref{rpa}) is the 
paramagnetic susceptibility of the Hartree-Fock approximation.
When we calculate $\chi^{(0)}$ of the usual single band Hubbard model, 
the Hartree-Fock terms
just renormalize the chemical potential. In our model
the Hartree term similarly changes the diagonal part of ${\cal H}_0$,
however the Fock terms introduce off-diagonal terms of the
orbital hoppings on the same site, giving rise to the crystal field effect.
This crystal field is totally symmetric concerning the entire unit cell,
but its local symmetry on each sublattice is trigonal.
The effect of this additional field is included in the
band-structure calculations. Therefore the free paramagnetic 
susceptibility calculated by using the tight-binding parameters
gives the appropriate $\chi_{nn'}^{\beta\beta'(0)}$.
In the actual calculation, we used the free $\chi^{(0)}$ per
tetrahedron of the form
\begin{eqnarray}
\chi^{(0)}_{\mu_1\mu_2\mu_3\mu_4}(q,\omega)&=&
\frac{1}{4L^3}\sum_k\sum_{\nu_1\nu_2}
\frac{f(\epsilon_{k+q\nu_2})-f(\epsilon_{k\nu_1})}
{\epsilon_{k\nu_1}-\epsilon_{k+q\nu_2}+\hbar(\omega+i\delta)}\nonumber\\
&&\hspace*{-2cm}\times U_{\mu_1\nu_1}(k)U_{\mu_4\nu_1}(k)
U_{\mu_2\nu_2}(k+q)U_{\mu_3\nu_2}(k+q),\label{kai0}
\end{eqnarray}
and rotated the orbital basis from $\{m_1m_2m_3m_4\}$ into 
$\{\beta\beta'\}$, where $f(\epsilon)$ is the Fermi distribution function.

\section{estimation of the specific heat coefficient}
Since we are primarily interested in the mass enhancement in $\gamma$,
let us now focus on the $\omega$-linear part of $Im\chi$ in the Free energy
of Eq. (\ref{F}). As is seen from the band structure (Fig. \ref{band}),
nothing singular would happen in $\chi^{(0)}$, therefore a constant
$Re\chi^{(0)}$ and an $\omega$-linear $Im\chi^{(0)}$ behaviors are expected
in the leading order of $\omega$. Then it is sufficient to
solve the RPA equations in the lowest order of 
$Im\chi^{(0)}\propto \omega$.
For the 1/2-filled $s$-band pyrochlore Hubbard model, 
in contrast, the complete matching of the Fermi level and flat 
bands brings a non-Fermi liquid behavior\cite{isoda,fujimoto2}.
As for the still-unfixed ${a_{\gamma\gamma}}$, we take
\begin{eqnarray}
a_{\gamma\gamma}=-\frac{1}{2}\Lambda_{\gamma\gamma}\equiv
-\frac{1}{2}
\left(\Lambda^{(c)}\oplus\Lambda^{(s)}\oplus\Lambda^{(s)}\oplus\Lambda^{(s)}
\right).
\end{eqnarray}
It is easy to check that the $a_{\gamma\gamma}$'s of this form
reproduce the original interaction term, Eq. (\ref{xx}). 
This choice is natural in the sense that
$\Lambda_{\gamma\gamma}$ reflects the symmetry of the non-interaction term
(SU(2) and $T_d$).
We neglect the weak dependence of $\chi^{(0)}$ on the interaction
through the Fock terms for simplicity, then
the $x$-integral in Eq. (\ref{F}) can be performed analytically. 
After some calculations, $\gamma$ coefficient (=$-\partial^2F/\partial^2T$)
per tetrahedron is given by 
\begin{eqnarray}
\gamma_{\rm RPA}
=\sum_{\beta=0}^{8}\left(\gamma^{\beta(c)}+3\gamma^{\beta(s)}\right),
\end{eqnarray}with
\begin{eqnarray}
\gamma^{\beta(c/s)}
&=&\frac{\pi k_B^2}{12L^3}\sum_{qn}\sum_{n_1\beta_1}
\Lambda^{(c/s)}_{\beta\beta}\frac{Im \chi_{nn_1}^{\beta\beta_1(0)}
(\vec{q},\omega)}{\omega}\nonumber\\
&&\times\left(
{\bf 1} -2\Lambda^{(c/s)}_{\beta_1\beta_1}Re\chi_{n_1n}^{\beta_1\beta(0)}
(\vec{q},\omega)
\right)^{-1}.
\end{eqnarray}
This equation is exact after taking $\omega,\delta,T\rightarrow 0$ 
and $L\rightarrow \infty$ limit. Note that 
we neglect the zero-point fluctuation term in the $\omega$ integral
because it 
will only show a weak temperature dependence through that of $\chi^{(0)}$.

\begin{figure}[hbt]
\includegraphics[width=86mm]{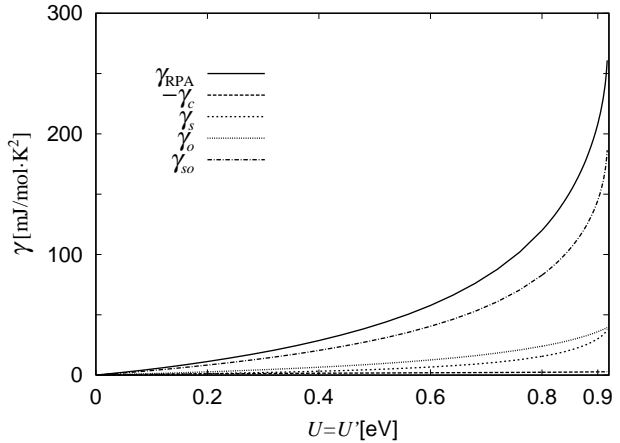}
\caption{\label{gamma} 
The graph shows $\gamma$ value per formula unit of ${\rm LiV_2O_4}$ mole. 
The contributions from charge, spin, orbital, and spin-orbital 
fluctuations are labeled by 
$\gamma_c, \gamma_s, \gamma_o$ and $\gamma_{so}$, respectively.}
\end{figure}

Finally we numerically estimate $\gamma_{\rm RPA}$ in the SU(6) limit
($U=U'$ and $J=0$), where the orbital fluctuations are mostly enhanced.
For this purpose, we calculated the free susceptibility, Eq. (\ref{kai0}),
for the parameter set of
$(\omega,\delta,T)={\rm (0.02eV,0.02eV,0.02eV)}$ and $L=16$.
We checked the convergence of $\chi^{(0)}$ against $\omega$ and $L$ 
along some symmetric lines.
Especially, 
we confirmed that $Im\chi ^{(0)}$ shows a clear $\omega$-linear dependence
in the range of $\omega=0.01, 0.02, 0.04, 0.08$ eV.
The convergence of the $\omega$-linear coefficient regarding
the system size ($L=12, 16, 24, 32$ ) is fairly good except for 
a few points very close to the ${\mit \Gamma}$ point.
As a result, we obtained the $\gamma$-value per ${\rm LiV_2O_4}$ mole
as shown in Fig. \ref{gamma}. The total $\gamma$ is the summation of
the contributions from charge (denoted by $\gamma_c=\gamma^{0(c)}$),
spin ($\gamma_s=3\gamma^{0(s)}$), orbital 
($\gamma_o=\sum_{\beta=1}^{8}\gamma^{\beta(c)}$), and spin-orbital fluctuations
($\gamma_o=3\sum_{\beta=1}^{8}\gamma^{\beta(s)}$).
The critical value $U_c$ is about 0.92 eV and there the complicated
eigenvectors, made of the four-sublattice summations of the spin and
spin-orbital components with quadrupole orbital modes, 
characterize the instability at the wave vector near 
${\mit \Gamma}$ point. 
Since we fixed $\omega$ finite and took the limit of $\vec{q}\rightarrow 0$, 
our calculated $Re\chi^{(0)}$ does not correspond to
the $\omega\rightarrow 0$ limit at ${\mit \Gamma}$ point.
Therefore we can not determine that the most prominent modes are just at
the ${\mit \Gamma}$ point or not. 
The real part of $\chi^{(0)}$ is studied for different hopping
parameters at zero temperature\cite{tsunetsugu}. 
In the present calculation, we do not detect
the peak structure around $X$ and $L_2$ points found in \cite{tsunetsugu}.
The reason for the difference might be that our calculations 
have done at finite $T$ and/or these single modes are quite sensitive 
to the nesting of the Fermi surface.

Fig. \ref{gamma} shows that
the correct order of $\gamma$ can be reproduced even away from
the critical point, say $\sim 0.8 U_c$. Around there,
the enhancement of $\gamma_{\rm RPA}$ (about 80\%)
is mainly attributed to the various spin-orbital fluctuations. 
Both spin and spin-orbital fluctuations contribute $\gamma_{\rm RPA}$
by the same order as a single component. However,
the magnitude of the components, 3 for spin and 24 for spin-orbital modes,
results in the different contribution in $\gamma_{\rm RPA}$.
In order to gain the same enhancement solely by
the spin fluctuations, $U$ must be very close to $U_c$ even
after the inclusion of $J$ term.
Away from the $U_c$, the enhancement of $\gamma_{\rm RPA}$ 
comes from the entire momentum space 
because the spin and spin-orbital fluctuations have rather 
weak $q$-dependences.
This result seems to be consistent with the naive picture that
the geometric frustration induces local fluctuations, which is
responsible for the heavy mass.

\section{conclusion}
In conclusion we have studied the origin of the large enhancement
of $\gamma$ in LiV$_2$O$_4$ based on the 1/4 filled $t_{2g}$-band 
pyrochlore Hubbard model. 
We introduced $T_d$ irreducible operators for orbital density operators
and derived the RPA equations in a simple form.
The $\gamma_{\rm RPA}$ obtained by the coupling constant integration
of $Im\chi$ is enhanced typically by order of 10 due to the spin-orbital
fluctuations compared with the $\gamma_s$ due to the spin alone. 
This conclusion itself
seems to be general for the orbital disordered system and thus
it may be possible to apply the present scenario to other systems,
such as Y(Sc)Mn$_2$ and $\beta$-Mn. 
To be more specific to LiV$_2$O$_4$, the calculated $\gamma_{\rm RPA}$ value
is of the same order as $\gamma_{\rm exp}$. 
However, it should be noted that our simple theory does not include the 
effect of the geometrical frustrations sufficiently, since the mode-mode
coupling between various fluctuations is neglected.
In general, the RPA approximation overestimates
the fluctuations, therefore the most prominent mode itself would be
suppressed and the $U_c$ becomes larger if we include the
couplings between the modes with different momenta.
At the same time, the overall structure of $\chi$ in $q$ space would
change into more and more structure-less and thus the additional enhancement 
of $\gamma$ could be expected.

It is well-known that the mode-mode coupling theory leads to the Curie-Weiss 
behavior of the magnetic susceptibility. As for the $R_{\rm W}$, it is
of order of 0.1 in the RPA approximation. This value would be 
increased by including $J$, although it is not clear whether the Hund's rule
coupling is sufficient to obtain $R_{\rm W} \sim 1.7$.
To address the similarity to the 4$f$ HF compounds, we need to develop
a more sophisticated theory including the coupling between fluctuations.
The fluctuation exchange approximation for the multi-band model
would be useful as the next step. 

\section*{ACKNOWLEDGMENTS}
We thank H. Tsunetsugu for sending us his related work\cite{tsunetsugu}
prior to the publication. 
Y.Y. is supported by the Japan Society for the Promotion of Science (JSPS).
K.U. is supported by Grant-in-Aid for Scientific Research from the JSPS.


\begin{thebibliography}{99}
\bibitem{kondo}S. Kondo $et\,al.$, Phys. Rev. Lett. {\bf 78}, 3729 (1997).
\bibitem{urano}C. Urano, M. Nohara, S. Kondo, F. Sakai, H. Takagi, 
T. Shiraki, and T. Okubo, Phys. Rev. Lett. {\bf 85}, 1052 (2000).
\bibitem{urano2}C. Urano, private communication
\bibitem{ueda}Y. Ueda, N. Fujiwara, and H. Yasuoka,
{\bf 66}, 778 (1997).
\bibitem{lee}S. -H. Lee, Y. Qiu, C. Broholm, Y. Ueda, and J. J. Rush,
Phys. Rev. Lett. {\bf 86}, 5554 (2001).
\bibitem{shiga} M. Shiga, K. Fujisawa, and H. Wada, 
J. Phys. Soc. Japan, {\bf 62}, 1329 (1993).
\bibitem{nakamura} H. Nakamura, K. Yoshimoto, M. Shiga, M. Nishi and K. Kakurai, 
J. Phys.: Condens. Matter {\bf 9}, 4701 (1997).
\bibitem{anisimov}V. I. Anisimov, M. A. Korotin, M. Z\"{o}lfl, T. Pruschke,
K. Le Hur, and T. M. Rice, Phys. Rev. Lett. {\bf 83}, 364 (1999).
\bibitem{matsuno}J. Matsuno, A. Fujimori, and L. F. Mattheiss,
Phys. Rev. B {\bf 60}, 1607 (1999).
\bibitem{eyert}V. Eyert, K. -H. H\"{o}ck, S. Horn, A. Loidl, and 
P. S. Riseborough, Europhys. Lett. {\bf 46}, 762 (1999).
\bibitem{singh}D. J. Singh, P. Blaha, K. Schwarz, and I. I. Mazin,
\bibitem{kusunose} H. Kusunose, S. Yotsuhashi, and K. Miyake,
Phys. Rev. B {\bf 62}, 4403 (2000).
\bibitem{flude} P. Flude, A. N. Yaresko, A. A. Zvyagin, and Y. Grin,
Europhys. Lett. {\bf 54}, 779 (2001).
\bibitem{fujimoto} S. Fujimoto, Phys. Rev. B {\bf 65}, 155108 (2002).
\bibitem{lacroix} C. Lacroix, Can. J. Phys. {\bf 70}, 1469 (2002).
\bibitem{laad}M. Laad, L. Craco, and E. Muller-Hartmann, cond-mat/0202531.
\bibitem{hopkinson}J. Hopkinson and P. Coleman, cond-mat/0203288.
\bibitem{tsunetsugu} H. Tsunetsugu, J. Phys. Soc. Japan, {\bf 71}, 1844 (2002).
\bibitem{group}X. Hamermesh, {\it Group Theory}, (Addison-Wesley, Reading, MA, 1962).
\bibitem{isoda} M. Isoda and S. Mori, J. Phys. Soc. Japan, {\bf 69}, 1509 
(2000).
\bibitem{fujimoto2}S. Fujimoto, Phys. Rev.B {\bf 64}, 085102 (2001).
\end{thebibliography}
\end{document}